# On-Surface Decarboxylation Coupling Facilitated by Lock-to-Unlock Variation of Molecules upon the Reaction

Shaoshan Wang[1,6,#], Zhuo Li[1,6,#], Pengcheng Ding[1,3,#], Cristina Mattioli[2], Wujun Huang[4], Yang Wang[1], André Gourdon[2], Ye Sun[3], Mingshu Chen[4], Lev Kantorovich[5], Xueming Yang[6], Federico Rosei[7], and Miao Yu[1,6,*]

1. School of Chemistry and Chemical Engineering, Harbin Institute of Technology, Harbin 150001, China
2. CEMES-CNRS, Toulouse 31055, France
3. Condensed Matter Science and Technology Institute, Harbin Institute of Technology, Harbin 150001, China
4. Department of Chemistry, Xiamen University, Xiamen 361005, China
5. Department of Physics, King's College London, The Strand, London WC2R 2LS, U.K.
6. Dalian Institute of Chemical Physics, Chinese Academy of Sciences, Dalian 116023, China
7. INRS Centre for Energy, Materials and Telecommunications, Varennes (Quebec) J3X 1S2, Canada

\# These authors contributed equally to this work.

Correspondence to: miaoyu_che@hit.edu.cn

**Abstract**

On-surface synthesis (OSS) involving relatively high energy barriers remains challenging due to a typical dilemma: firm molecular anchor is required to prevent molecular desorption upon the reaction, whereas sufficient lateral mobility is crucial for subsequent coupling and assembly. By locking the molecular precursors on the substrate then unlocking them during the reaction, we present a strategy to address this challenge. High-yield synthesis based on well-defined decarboxylation, intermediate transition and hexamerization is demonstrated, resulting in an extended and ordered network exclusively composed of the newly-synthesized macrocyclic compound. Thanks to the steric hindrance of its maleimide group, we attain a preferential selection of the coupling. This work unlocks a promising path to enrich the reaction types and improve the coupling selectivity hence the structual homogeneity of the final product for OSS.

The development of nanosale organic semiconductors, such as macrocyclic aromatic hydrocarbons that possess conjugated π-electrons, is very promising for next-generation electronic devices, e.g. organic field-effect transistors, photovoltaics, and light-emitting devices.[1–3] Despite significant progress in producing these compounds in solution,[4–6] the sophisticated purification, the low solubility of the product as well as the difficulty in evaporating them onto substrates and controlling the growth of molecular films are major unresolved challenges that hinder their use in applications. In this framework, on-surface synthesis (OSS) via triggering the reactions of molecular precursors directly deposited on suitable substrates is a consolidated strategy with huge potential. [7–16]

However, unlike liquid phase synthesis which is compatible with numerous reaction types, OSS has been mainly confined to a few specific reactions which typically have a relatively low reaction temperature, e.g. Ullmann coupling,[10,11] Glaser coupling,[12,13] Bergman cyclization,[14] and C–H bond activation.[15,16] Synthesis processes requiring a relatively high reaction temperature are less practicable for OSS. For example, decarboxylation is an essential process for eliminating carboxylic acid in petroleum or coal,[17] and a critical reaction type in synthetic chemistry.[18–20] Nevertheless, when applying this type of reaction on surfaces, both the reaction yield and the resultant molecular coverage are low;[21–24] the final products are poorly ordered and often mixed with the unreacted precursors and/or intermediates. When relatively high energy barriers are involved, OSS approaches faces a primary issue which is related to a typical dilemma: a firmly anchored adsorption is necessary for maintaining the molecules on the surface sufficiently long to overcome the reaction barrier;[25,26] on the other hand, the molecular mobility is in itself important for the subsequent molecular arrangement and coupling.[27,28] In addition, due to the known competition between the chain and ring coupling forms on surfaces,[13,29–32] another issue for OSS relates to the difficulty in controlling the structural homogeneity of the final product.

Here, we present a strategy to address the above-mentioned challenge of OSS, using a specially-designed precursor, 3,5-bis(carboxyl acid)-phenyl-3-maleimide ($C_{12}H_7NO_6$, BCPM, Scheme 1).[33] Its maleimide group can be firmly locked on Cu(111) when BCPM is intact and unlocked after decarboxylation, and also provides the necessary steric hindrance to enable the selectivity of the specific coupling form. Combining scanning tunneling microscopy (STM), infrared reflection-absorption spectroscopy (IRAS) and density functional theory (DFT) calculations, we demonstrate high-yield synthesis of a macrocyclic compound based on decarboxylation, intermediate transition and hexamerization, with a well-maintained molecular coverage throughout the reactions. The resultant ordered and extended molecular adlayer is exclusively composed of the newly-synthesized hexamer rings, whereas chain-like coupling form is hindered. Our work thus defines a special opportunity to address the challenges related to high reaction energy barrier and the structual homogeneity for OSS.

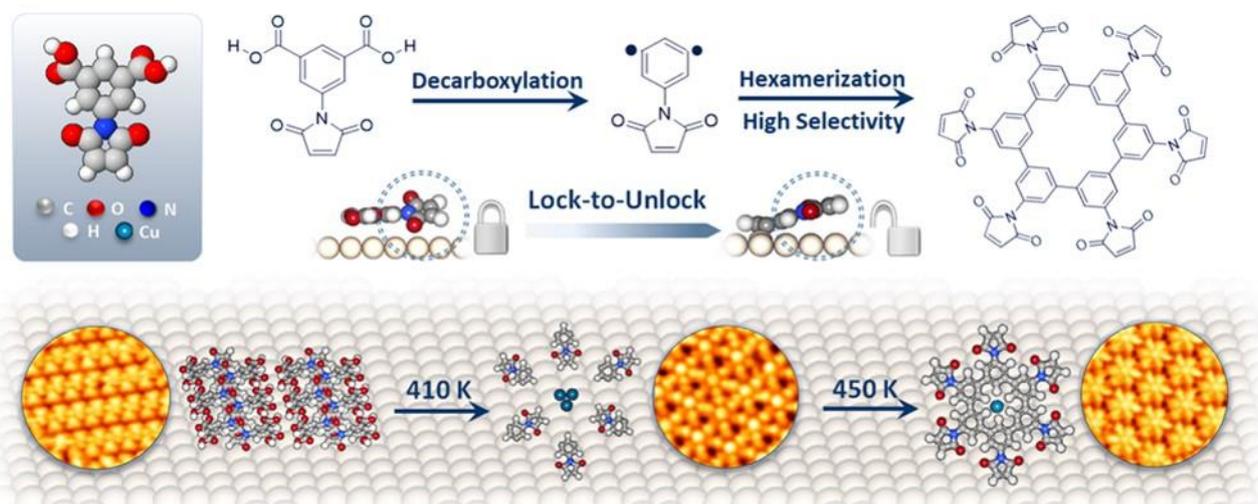

**Scheme 1.** Illustration of BCPM compound and its on-surface decarboxylation, intermediate transition and hexamerization on Cu(111).

After depositing BCPM onto Cu(111) kept at room temperature (RT) in ultrahigh vacuum, the molecules self-assemble into a distinct, ordered structure (Figure 1a–b), with parallel zipper-like chains packed side-by-side into extended domains (Figure 1a and S1). Each zipper is composed of two interlocked rows. Each 'zipper teeth' is assigned to an individual BCPM molecule, where the circular and rounded-triangular protrusions correspond to the maleimide and *bis*(carboxyl acid)- phenyl group, respectively (see the inset in Figure 1a). The periodicity of molecules along the rows is 10.2±0.2 Å, and the center-to-center distance between the nearest-neighboring zippers is 15.8±0.3 Å.

This '*zipper*' structure was then explored by DFT calculations. As shown in Figure 1c, for each BCPM in the network, the phenyl ring and two carboxylic acid groups adopt a flat geometry at an apparent height of 3.1 Å above the substrate, whereas the maleimide ring is rotated by 35° from the horizontal direction. In this form, one side of the maleimide ring is closer to the surface than the other, and the lower carbonyl oxygen ($O_L$) of the maleimide sits exactly on top of a Cu atom. The $O_L$–Cu distance is 2.3 Å. All the molecules in the '*zipper*' structure are found to adopt this adsorption conformation (Figure 1d). The molecular adlayer has base vectors (green arrows) $|a|$=10.2 Å, $|b|$=15.6 Å. The simulated STM image of the molecular network (Figure 1e) shows the morphology and molecular arrangement consistent with the experimental observation. The maleimide ring of an isolated single BCPM is also tilted on Cu(111) in the same manner as that in the 'zipper' structure (Figure S2); the maleimide ring rotation is therefore not induced by the close packing of BCPMs. The charge density difference plot reveals evident electron donation from Cu to $O_L$ (Figure 1f). Based on Bader charge analysis [details in Supporting Information (SI)], the Cu-to-$O_L$ charge transfer is 0.14 *e*. The mapping of the electron localization function (ELF, Figure 1g) demonstrates that the localized charge is distributed near $O_L$, confirming an ionic bonding between $O_L$ and the Cu substrate.[34] As a result, the adsorption energy of BCPM on Cu(111) in this structural phase is –1.79 eV per molecule, and the intermolecular interaction is –0.51 eV per molecule.

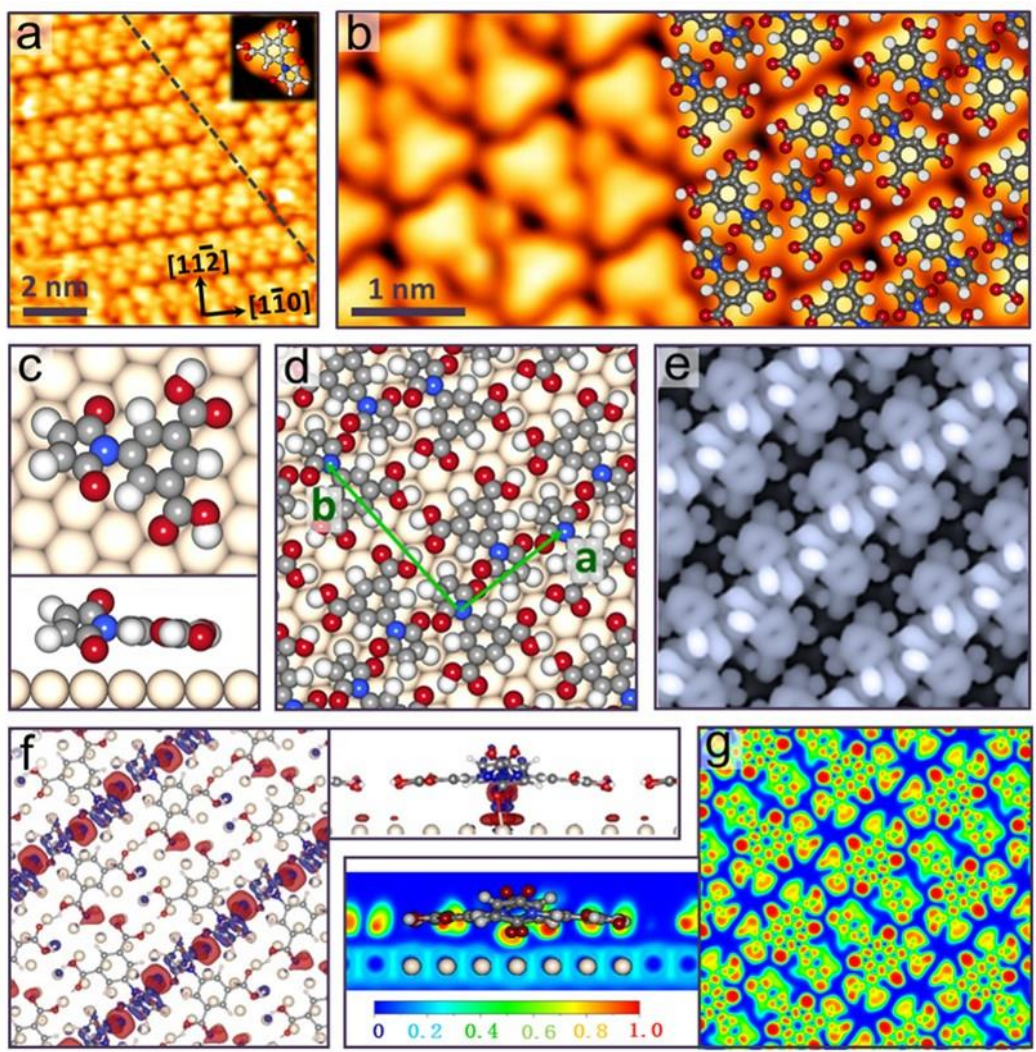

***Figure 1.*** (a–b) STM images of the BCPM domain formed on Cu(111) at RT without post treatment, where each zipper-tooth-like motif is assigned to an individual molecule. (c–d) DFT optimized model of the '*zipper*' structure, where the top and side views of an individual BCPM molecule in the network is presented in c. (e) Simulated STM image based on the structural model in d. (f) The top and side views of the charge density difference with the isosurface value being 0.01 e·Å$^{-3}$, and the red and blue colors indicating charge depletion and accumulation, respectively. (g) ELF mapping for the extended '*zipper*' structure.

To enable decarboxylation, the sample is annealed at progressively increasing temperatures. After annealing at ~350 K, even though most portions of molecular domains keep the 'zipper' structure, randomly distributed molecular motifs are observed at the domain's edges (Figure S3). Further annealing to ~370 K leads to higher percentages of disordered areas. When the temperature increases to ~400 K, an ordered transition phase starts to emerge, denoted as 'porous network (PN)'. Upon further annealing at 410~430 K, the 'PN' phase becomes dominant (Figure 2a). As marked by the two arrows in Figure 2b, its unit cell has lateral sizes of |a|=|b|=22.0±0.5 Å, containing three long ellipsoids (oriented by 120° relative to one another) together with one circular protrusion. Each long ellipsoid is also imaged as two sub-protrusions, with the same long-axis length as that of the BCPM molecular board. However, the profile corresponding to the bis(carboxyl acid)-phenyl group is not a triangle any more, as it appears more rounded.

To understand the 'PN' structure, we carried out DFT calculations. The molecular motif is attributed to that of decarboxylated BCPM with two radicals (PM-2R, Figure S4, details in SI). As shown in Figure 2c, the three PM-2R in each unit cell are oriented 120° relative to one another. For each PM-2R (Figure 2d), the phenyl ring is bent towards the substrate; two substrate Cu atoms are lifted up (0.33 Å) by their interaction with the decarboxylated phenyl, leading to a C–Cu length of 2.0 Å; the two O atoms in the maleimide are at the same height above the substrate (3.3 Å). They are no longer in the "lock-in" geometry as in the case of intact BCPM. Each circular protrusion in Figure 2a–b is attributed to a Cu cluster containing three Cu atoms, as its size is significantly larger than a single Cu atom and $CO_2$ produced by decarboxylation is known to easily desorb from the surface (desorption temperature ~80 K).[35] Each Cu atom of the cluster is located on the Cu(111) hollow sites at an apparent height of 2.1 Å above the outermost Cu layer. The calculated STM image (Figure 2e) shows that the dimension and morphology of the circular protrusion and molecular motifs as well as lateral geometry of the network are consistent with the experimental results. According to the charge density difference plot (Figure 2f) and ELF mapping (Figure 2g), the phenyl ring of PM-2R is strongly bonded with the two Cu substrate atoms with a Cu-to-C charge transfer of 0.16 $e$, whilst the maleimide ring shows no charge transfer with the substrate. The adsorption energy of each PM-2R is –4.98 eV, primarily attributed to the interaction between Cu and the decarboxylated phenyl group.

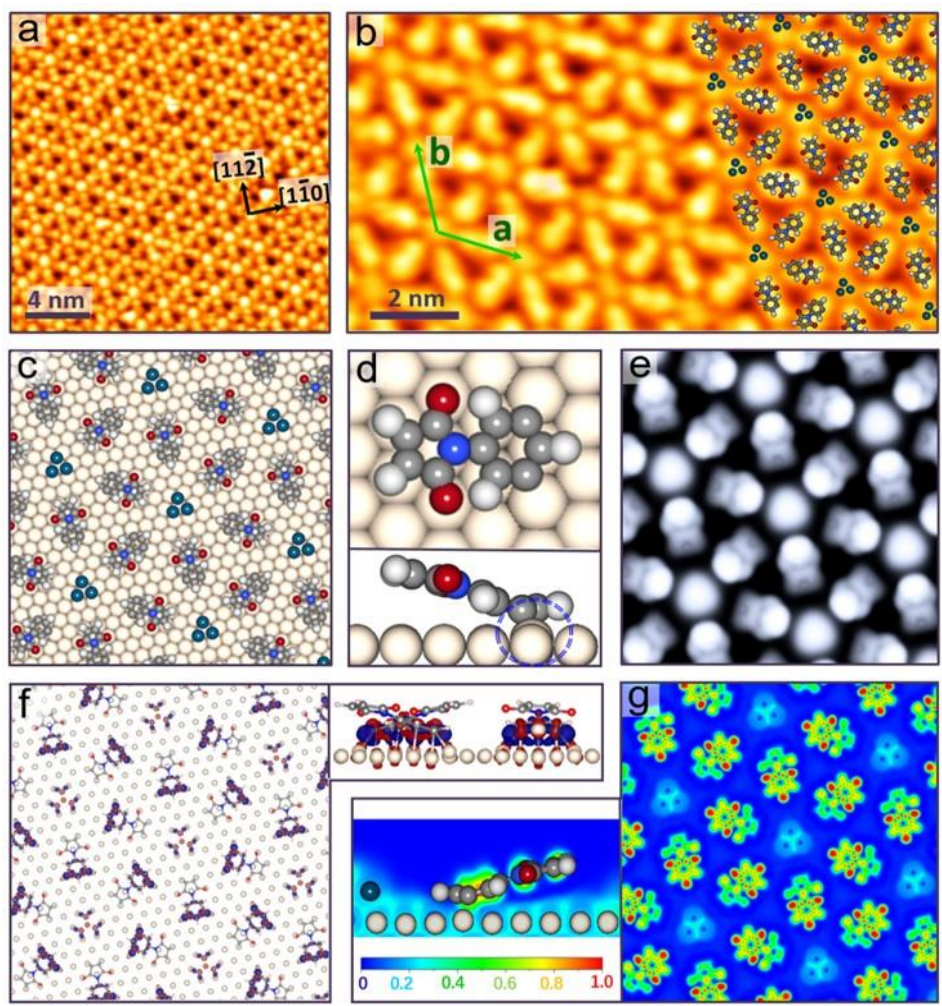

*Figure 2.* (a–b) STM images of the '*PN*' intermediate phase, where the two lattice vectors are marked by green arrows. (c) The DFT calculated model for the extended network of PM-2R and Cu clusters, well commensurate with the substrate. (d) The top and side views of DFT optimized single PM-2R on Cu(111), where the substrate Cu atoms are lifted up (blue circle) due to the strong interaction with the decarboxylative phenyl and the maleimide group is not rotated to lower one of its O atoms any more. (e) Simulated STM image, the top and side views of (f) the charge density difference and (g) ELF mapping for the '*PN*' structure. In f, the isosurface value is 0.045 e·Å$^{-3}$.

Upon further annealing to ~450 K, the 'PN' structure gradually shrinks until it vanishes, and flower-like species randomly emerge on the substrate (Figure S5). After annealing at ~470 K for 20 min, extended and well-defined domains tiled by the 'flower' species are fabricated (Figure 3a); its Fourier transform (FT) pattern is presented in the inset. When annealing at higher temperatures (> 500 K), the adlayer coverage starts to reduce with temperature until full desorption.

The ordered arrangement of the 'flowers' adopts a three-fold symmetry, with its close-packed direction along the <11-2> direction of the substrate and the unit cell dimension of |a|=|b|=22.0±0.5 Å. Each 'flower' motif is composed of six petal-like protrusions and a circular lobe in the flower center (Figure 3b), with each petal orientated by ⁓12⁰ relative to the <1-10> direction of the Cu(111) surface. According to the molecular size and morphology revealed by the STM images, the flower motif corresponds to a new compound formed via hexamerization of PM-2R, i.e. six phenyl-3-maleimide (6-PM) bonded to one another with their phenyl rings oriented inward. A single Cu adatom sits in each flower's center (Figure S6, details in SI).

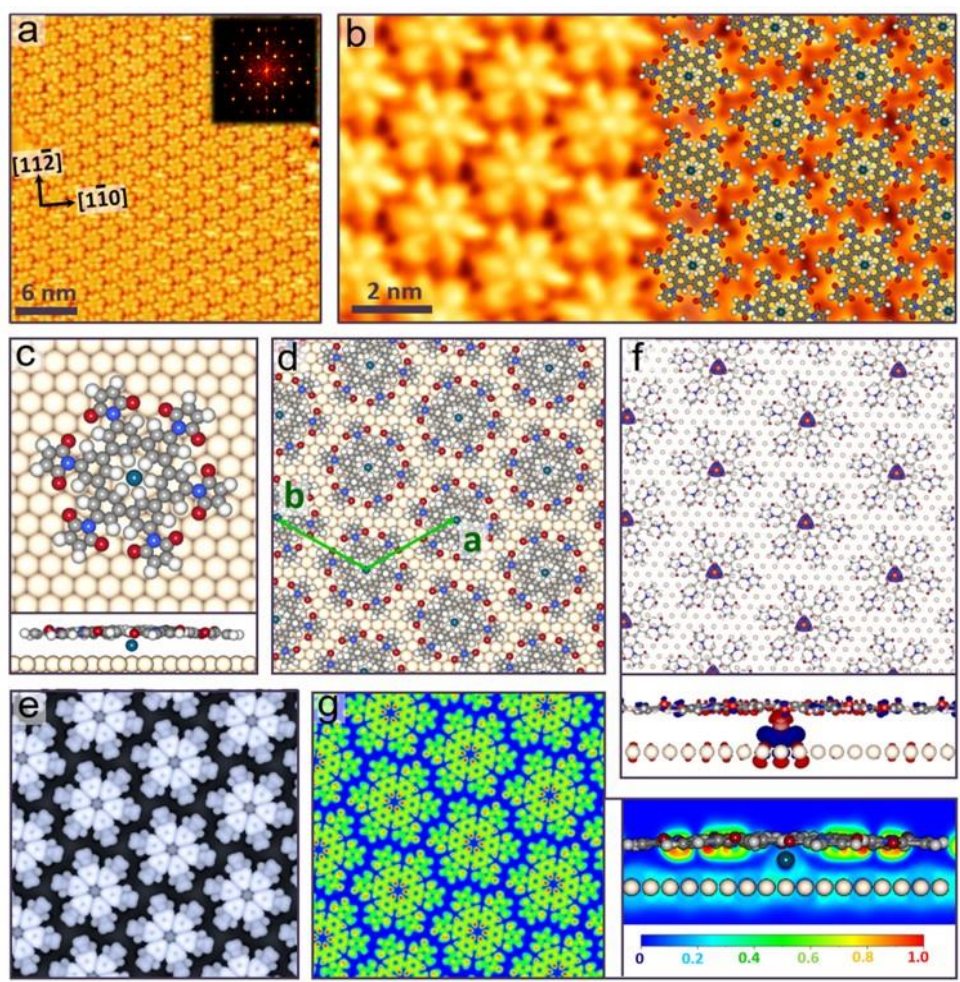

*Figure 3.* (a–b) STM images of the extended network composed of flower-like species on Cu(111), where the FT pattern derived from the STM image is presented in the inset of a. (c) Top and side views of the DFT optimized model for an individual 6-PM molecule in the flower network on Cu(111). (d) DFT calculated model and (e) simulated STM image of the '*flower*' network of 6-PM, in good agreement with the experimental observations. (f) Top- and side-view plots of the charge density difference and (g) ELF mapping of the '*flower*' structure. The isosurface value is 0.012 e Å$^{-3}$ in f.

The '*flower*' network of 6-PM was further investigated by DFT calculations. As shown in Figure 3c, 6-PM is adsorbed in a nearly flat conformation at a height of 3.1 Å above the Cu(111) surface: the six phenyl rings are all parallel to the substrate and the maleimide groups adopt a nearly flat form (5.5° tilt relative to the surface); the central Cu adatom adsorbs at a height of 2.1 Å above the substrate, sitting on the three-fold hollow site of Cu(111). This extended network is commensurate with the substrate, adopting a (5√3×5√3) R30° structure (Figure 3d). The molecular morphology and lateral arrangement of the simulated image (Figure 3e) are in good agreement with the experimental observations. The charge density difference plot (Figure 3f) reveals no charge transfer between 6-PM and the substrate; there is evident charge transfer from the central Cu adatom to the substrate, resulting in a positive charge of +0.17 $e$ on the central Cu adatom. ELF mapping (Figure 3g) further confirms no charge localization on 6-PM. The adsorption energy for each 6-PM '*flower*' is –5.34 eV, *i.e.* –0.89 eV per PM.

To better explore the decarboxylation reaction, we performanced *in situ* IRAS analysis. Compared with the spectrum collected at RT, the IR absorption peaks of the sample show an obvious variation in the 400–490 K range (Figure 4). The signal at 557 cm$^{-1}$ is attributed to the stretching vibration of C–Cu;[36,37] it emerges at ~410 K, reaches the maximum intensity at ~430 K, decreases gradually at higher temperatures, and vanishes at ~470 K. This variation of C–Cu bonding supports the occurrence of decarboxylation and the strong bonding between the decarboxylated phenyl and the lifted Cu substrate atoms. The distinct peak at 1420 cm$^{-1}$ is attributed to the in-plane symmetrical stretching vibration of O=C–O.[38,39] The significant decrease of this peak at ≥ 410 K also provides direct evidence for decarboxylation. The broad peak at 1732 cm$^{-1}$ corresponds to the stretching vibration of C=O,[38,39] and the absorption frequency of C=O in the carboxylic group is higher than that in the maleimide ring.[40] Upon annealing at ≥ 410 K, this peak shifts from 1732 cm$^{-1}$ to 1728 cm$^{-1}$, with a reduced integral intensity. These variations further confirm the occurrence of a decarboxylation reaction, as only the signal corresponding to C=O in the maleimide ring remains after the decarboxylation.

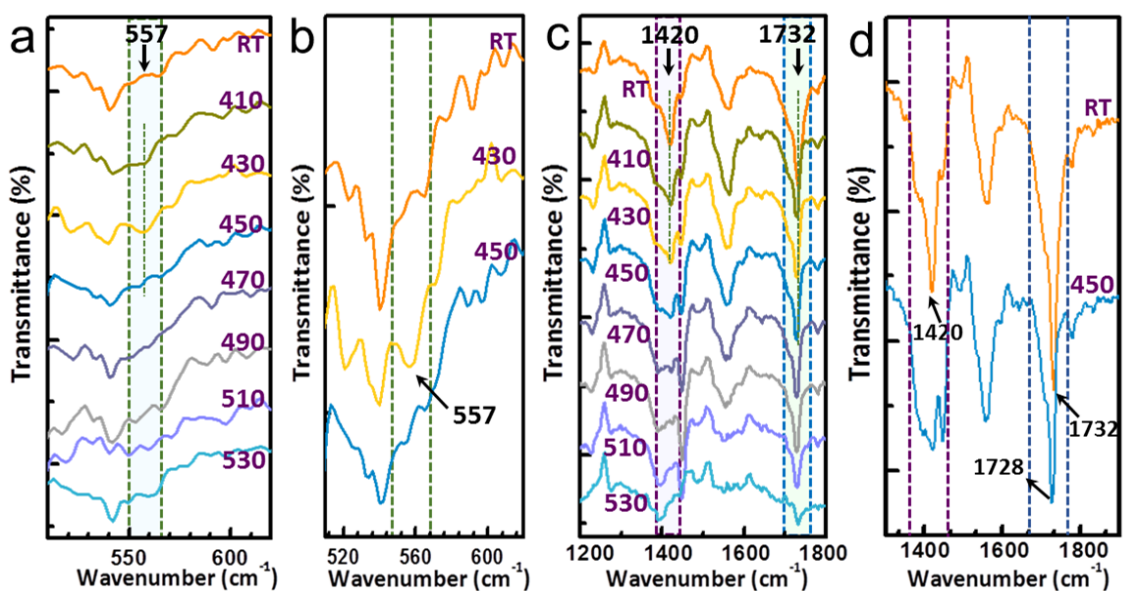

*Figure 4. In situ* IRAS spectra of the BCPM adlayer on Cu(111) when annealing at gradually increased temperature with the vibration wavelength region in (a–b) 510–620 cm$^{-1}$, (c) 1200–1800 cm$^{-1}$, and (d) 1300–1900 cm$^{-1}$. Compared with the spectrum collected at RT, a new peak (557 cm$^{-1}$) appears at 410 K, increases with the temperature and vanishes at 470 K; the peak at 1420 cm$^{-1}$ significantly decreases at ≥410 K; the peak at 1732 cm$^{-1}$ shifts to 1728 cm$^{-1}$ with a reduced integral intensity. These variations are consistent with BCPM decarboxylation.

To understand the reaction path, we conducted climbing image nudged elastic band (CI-NEB) calculations. To lower the computational cost, a simpler molecule, *i.e.* benzoic acid, was employed as precursor instead of BCPM (Figure S7). The calculated energy barriers for the decarboxylation and coupling between phenyl rings are 1.15 eV and 0.42 eV, respectively. The coupling undergoes an exothermic path (−1.97 eV).

Compared with previously reported OSS based on decarboxylation coupling,[21–24] our case is distinct: the molecules barely desorb during the whole reaction process; the molecular arrangement and reaction are independent of molecular coverage; the decarboxylation step can be triggered at a lower temperature; the reaction yield is high (>85% according to the statistical analysis of STM results); the final molecular adlayer exhibits long-range order; the hexamer ring structure appears as the only product, with chain-like coupling completely hindered.

The merits are related to the varied adsorption conformation upon decarboxylation and steric hindrance effect of the maleimide group. The maleimide ring in intact BCPM adopts the tilted form on Cu(111), locking the molecules by strong Cu-$O_L$ ionic bonding. This firm locking contributes to the well-maintained molecular coverage over decarboxylation, and induces the same assembly structure and reaction form for largely varied molecular coverages. Moreover, such firm molecular bonding on the surface also favors the reaction yield: if molecules diffuse and stay only a short time at stable geometries from which the reaction/transition is possible, the overall probability of the reaction is reduced. The maleimide ring becomes 'unlocked' upon decarboxylation, facilitating the following coupling and long-range order of the synthesized molecular network. The 'lock-to-unlock' variation of maleimide is closely associated with the intramolecular electron distribution (Figure S8). Based on the electrostatic potential-fitted charges, for intact BCPM, due to the high electron affinity of the carboxylic groups, the average charges on the two O atoms of maleimide are –0.40 $e$ and 0.41 $e$, respectively. To accommodate the electron deficiency, the maleimide group rotates so that its $O_L$ draws –0.14 $e$ from the underlying Cu atom, inducing a strong $O_L$–Cu interaction. After decarboxylation, the intramolecular charge is re-distributed to attain a new balance: each O atom of the maleimide group draws an additional –0.11 $e$ compared with that of the unreacted molecule; in this case, the need to gain extrinsic electron from the Cu substrate is eliminated so that the maleimide group rotates from the tilted form to a nearly flat geometry. In addition, the negatively charged O atoms of the maleimide group provide an electrostatic repulsion between neighbouring maleimide rings. Together with its steric hindrance, the hexamer ring appears to be the only coupling form, as the distance between two neighbouring O atoms in a zigzag chain-like coupling is too short to form (Figure S9).

In summary, we report a high-yield, long-range ordered synthesis of a large organic compound by on-surface decarboxylation coupling. The molecular locking enables high molecular stability for bond scission, leading to high yield, well-maintained molecular coverage, and lower reaction temperature; the subsequent unlocking facilitates the coupling and assembly of reacted species. The lock-to-unlock variation is associated with intramolecular electron re-distribution over the reaction. The steric hindrance effect of maleimide enables high selectivity to the coupling form. Our work opens up a promising route for improving the practicality of OSS in cases that involve high energy barriers as well as achieving the structural homogeneity and long-range order of the final product, which are major unresolved challenges in this field.

*Acknowledgements*

This work is financially supported by the National Natural Science Foundation of China (21473045, 51772066).

*Conflict of interest*

The authors declare no conflict of interest.

**Keywords:** on-surface synthesis • high energy barrier • decarboxylation coupling • intramolecular charge redistribution